\documentclass[twocolumn,superscriptaddress]{revtex4}
\usepackage{graphicx}
\usepackage{epsfig}
\usepackage{amsmath,amssymb}
\usepackage{amsfonts,amsbsy}
\usepackage{hyperref}
\usepackage{bm}
\usepackage{color}
\usepackage{enumerate}
\usepackage{verbatim}
\usepackage{lipsum}

\def\be{\begin{equation}}
\def\ee{\end{equation}}
\def\bea{\begin{eqnarray}}
\def\eea{\end{eqnarray}}

\begin{document}
\title{Rapidity window dependence of ridge correlations in the glasma}
\author{Donghai Zhang}
\affiliation{Key Laboratory of Quark and Lepton Physics (MOE) and Institute of Particle Physics, Central China Normal University, Wuhan 430079, China}
\author{Yeyin Zhao}
\affiliation{Key Laboratory of Quark and Lepton Physics (MOE) and Institute of Particle Physics, Central China Normal University, Wuhan 430079, China}
\author{Mingmei Xu}
\email{xumm@mail.ccnu.edu.cn}
\affiliation{Key Laboratory of Quark and Lepton Physics (MOE) and Institute of Particle Physics, Central China Normal University, Wuhan 430079, China}
\author{Xue Pan}
\affiliation{School of Electronic Engineering, Chengdu Technological University, Chengdu 611730, China}
\author{Yuanfang Wu}
\email{wuyf@mail.ccnu.edu.cn}
\affiliation{Key Laboratory of Quark and Lepton Physics (MOE) and Institute of Particle Physics, Central China Normal University, Wuhan 430079, China}
\date{\today}
\begin{abstract}
    We study ridge correlations of the glasma in pp collisions at $\sqrt{s_{\mathrm{NN}}}=7$ TeV by using the color glass condensate (CGC) formalism. The azimuthal collimation at long range rapidity is intrinsic to glasma dynamics and is reproduced here. When rapidity window enlarges, ridge correlations in two dimensional $\Delta y$-$\Delta\phi$ distribution and one dimensional $\Delta\phi$ distribution at long range rapidity gap are enhanced. The enhancements are demonstrated to be the contributions of source gluons. The quantum evolution of the gluons presents unique correlation patterns in differential correlation function. These characters of two gluon correlations open a way of testing the production mechanism from experimental measurements.
\end{abstract}

\maketitle
\section{Introduction}\label{sec:introduction}
The long range rapidity ridge correlation in small systems is one of the most unexpected discoveries~\cite{CMS-pp-JHEP-2010,CMS-pPb-PLB-2013,CMS-pPb-PbPb-PLB-2013,ATLAS-pPb-PRL-2013,ALICE-pPb-PLB-2013,ALICE-pPb-PLB-2016,ridge-RHIC-small-1,ridge-RHIC-small-2}. The structure of ridge in two dimensional $\Delta\eta$-$\Delta\phi$ correlations is twofold: a long ridge in $\Delta\eta$ direction and two peaks in $\Delta\phi$ direction. The two peaks at $\Delta\phi=0$ and $\pi$ are also termed the collimation production, revealing the collectivity at long range rapidity in small systems. The physics underlying the collectivity at long range rapidity is open yet and under extensive studies. 

Final state interactions, especially the hydrodynamic description, combined with a variety of initial conditions, attribute the long range ridge to the fluid nature~\cite{pp-hydro1,pp-hydro2,pp-hydro3,pp-hydro4,pp-hydro5,pp-hydro6}. The transverse expansion of the fluid driven by the pressure gradient can explain the azimuthal anisotropy of final state particles. The difficulties of hydrodynamic description are that it can not reproduce the multi-particle cumulant $c_2\{4\}$~\cite{hydro-cumu-1,hydro-cumu-2} and the elliptic flow of heavy flavor particles~\cite{hydro-flow-hf-1,hydro-flow-hf-2}. In these respects, CGC calculations reproduce the data well~\cite{cgc-c2,cgc-heavy}. Only initial correlations from CGC  can explain the long range ridge yield very well, too~\cite{cgc-PRD-I,cgc-PRD-II,cgc-PRD-III}. The trouble of CGC is that it can not generate the right ordering of Fourier harmonics $v_{n}$ in the three system sizes of pAu, dAu and HeAu~\cite{cgc-vn-ordering}. In a word, the two mechanisms, i.e. hydrodynamics and CGC, can only describe some of the data in small systems, respectively. Therefore, they both need further studies.

CGC describes the initial state of a colliding hadron or nucleus. The gluon density grows rapidly and gets saturated as the collision energy increases or as Feynman variable $x$ decreases. At saturation the small-$x$ gluons have large occupation numbers (of order $1/\alpha_s$), so their mutual interactions can be treated as the classical field with a strong field strength (of order $1/\sqrt{\alpha_{s}}$). Thus two colliding hadrons or nuclei at very high energies are two sheets of colored glass approaching one another. After the collision, strong longitudinal color electric and magnetic fields are formed in the region between the nuclei, which is called glasma. The glasma fields are localized in the transverse scale (of order $1/Q_s$) that are smaller than the nucleon size, which gives the picture of glasma flux tube. The strong longitudinal color fields in glasma flux tubes are approximately boost invariant and multi-particle productions in the small $x$ region naturally generate long range longitudinal correlations~\cite{cgc-NPA-2008,cgc-NPA-2010}. 

The correlations between gluons in high energy scatterings are caused by the quantum evolution, which is illustrated in Fig.~1(a). The short lines represent valence quarks, which radiate gluons in the form of gluon cascades. Besides the gluon radiation, when the gluon density gets large, recombinations of radiated gluons (non-linear effects) are also important and shown in the evolution. They result in the saturation phenomenon of the gluon distribution at $k_\perp\ll Q_s$, while the gluon distribution at $k_\perp\gg Q_s$ follows a behavior of linear approximation. Therefore, $Q_s$, called saturation momentum, as a scale separating linear and non-linear behavior, is the only parameter which can determine the physics in a given collision. This indicates an equivalence between nuclei and protons. The universality of hadronic interactions at high energies in the CGC effective field theory succeeds in explaining long range azimuthal correlations in pp, pPb systematically~\cite{cgc-PRD-I,cgc-PRD-II,cgc-PRD-III} and may account for the similarity between PbPb and pPb with the same multiplicity~\cite{Dusling-PRL-2018}. 

The CGC framework succeeds in explaining not only the multiplicity and transverse momentum dependence, but also the $\Delta\eta$ acceptance dependence of the long range azimuthal correlations in the LHC pPb data at 5.02 TeV~\cite{cgc-PRD-III}.  With a common set of parameters, ref~\cite{cgc-PRD-III} gets good fits to the CMS, ALICE and ATLAS data of different acceptances simultaneously.  The overall agreement with data is a spectacular achievement of CGC. Because they considers the combined effect of rapidity acceptance and normalizations of different LHC experiments, an explicit dependence on the rapidity acceptance alone is not presented yet. 

In high energy limit, both rapidity and transverse momentum of a gluon are related to its Feynman $x$. For a right moving projectile, the relation reads  
\begin{equation}x=\frac{\mathrm{p_{\perp}}}{\sqrt{s}}e^{y},\end{equation} with  $y$ representing rapidity, $\mathrm{p_{\perp}}$ transverse momentum and $\sqrt{s}$ center-of-mass energy. The dependence on rapidity is much more sensitive due to the exponential function. At $\sqrt{s}=7$ TeV and intermediate $\mathrm{p_{\perp}}$, e.g. 2 GeV$/$c, the rapidity region $y\leqslant 1.2$ corresponds to $x\lesssim 10^{-3}$, while $y\leqslant 3.5$ corresponds to $x\lesssim 10^{-2}$. It means that gluons at central rapidity region ($|y|\leqslant 1.2$) reflect properties of the small-$x$ ($x\lesssim10^{-3}$), as the green band shows in Fig.~1(b). In that rapidity region quantum evolutions are essential. In contrast, gluons at middle rapidity regions ($1.2< |y|\leqslant 3.5$), as the blue bands show in Fig.~1(b), present features of moderate-$x$ ($10^{-3}<x<10^{-2}$)  degree of freedom. Within this region, radiated gluons still dominate, but contributions of color sources begin to show up. The large-$x$ ($x\gtrsim10^{-2}$) degrees of freedom, i.e. gluons at large rapidity ($|y|> 3.5$) , act as sources and are referred to as source gluons, which are denoted by red bands in Fig.~1(b). 

It has been demonstrated that, the strong correlation between radiated gluons and source gluons can explain long range rapidity correlations~\cite{zhao-1, zhao-2}. However,  ridge correlations in experiments are usually measured within certain rapidity or pseudorapidity windows. For example, the pseudorapidity windows of ALICE and CMS are [-0.9, 0.9] and [-2.4, 2.4], respectively. A quantitative comparison of results from different experiments requires a study of the rapidity window dependence of ridge correlations.  

On the theoretical side, different rapidity windows are expected to reveal gluon dynamics at different $x$ values. Fig.~2 demonstrates that different rapidity windows ($Y_{\rm W}$ for short) pick pairs of different $x$ when a given rapidity gap like $\Delta y=2$ is being concerned. A narrow rapidity window only include contributions of radiated gluons, while a wider rapidity window can further include correlations between radiated gluons and source gluons. So, rapidity window dependence of long range ridge correlations is sensitive to the quantum evolution of gluon saturation dynamics.


Considering the above reasons, it would be valuable to study the rapidity window dependence of the ridge correlations systematically. This paper is organized as follows. In section \ref{sec:formula}, the definition of correlation function and some related formulae of single- and double-gluon inclusive production in CGC framework are given. The formulae in this manuscript follow those in Ref.~\cite{cgc-PRD-I,cgc-PRD-II,cgc-PRD-III} and are identical with those at gluon level without fragmentation functions. The new aspects of analysis method here lie in the exploration of contributions of different $x$ degrees of freedom. Results of correlations are shown and discussed in section \ref{sec:results}, where the sensitivity to rapidity windows are carefully compared. Section
\ref{sec:summary} gives the summary and discussion.

\begin{figure}[htbp]
\centering
\includegraphics[width=0.45\textwidth]{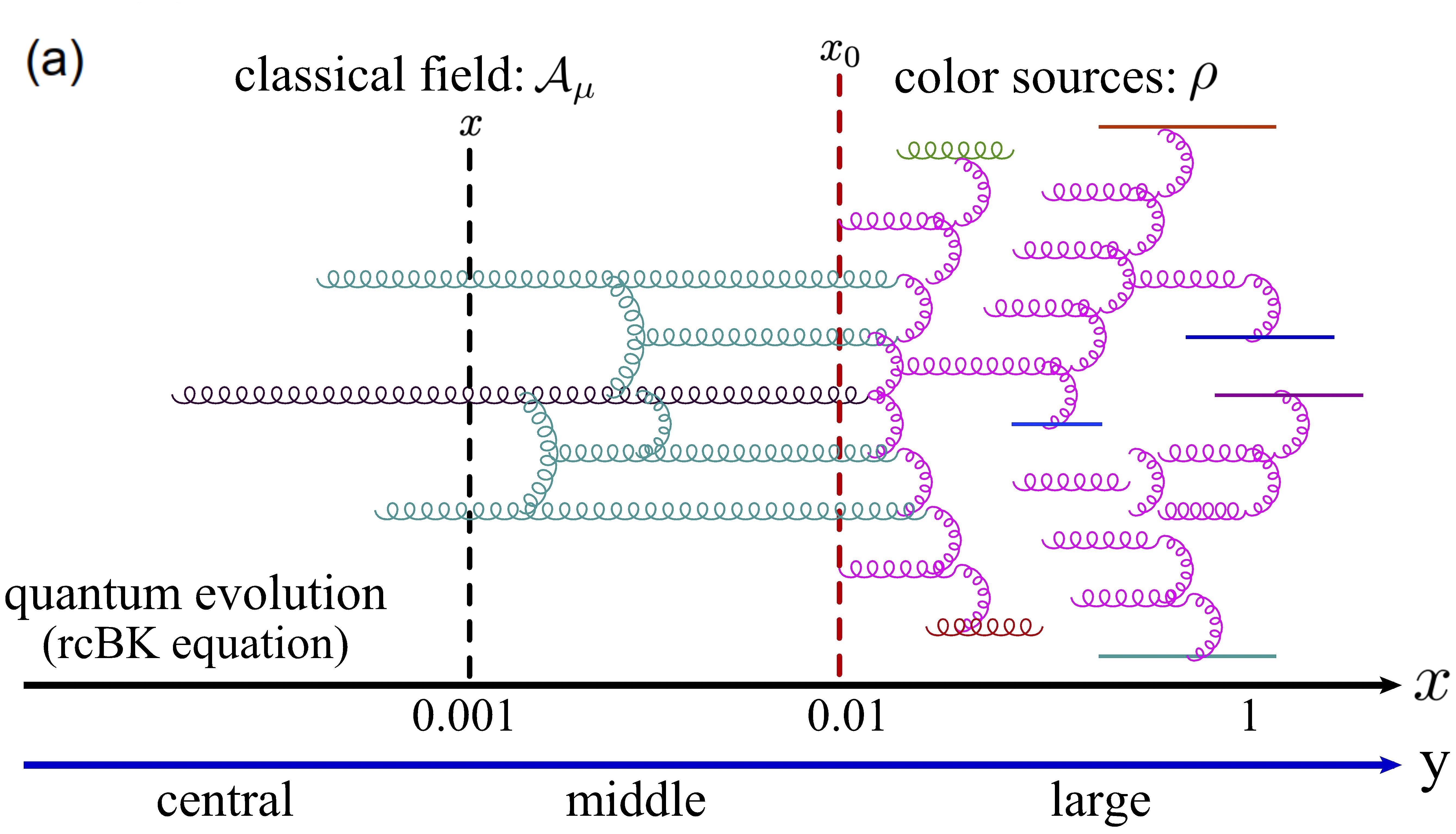}
\includegraphics[width=0.45\textwidth]{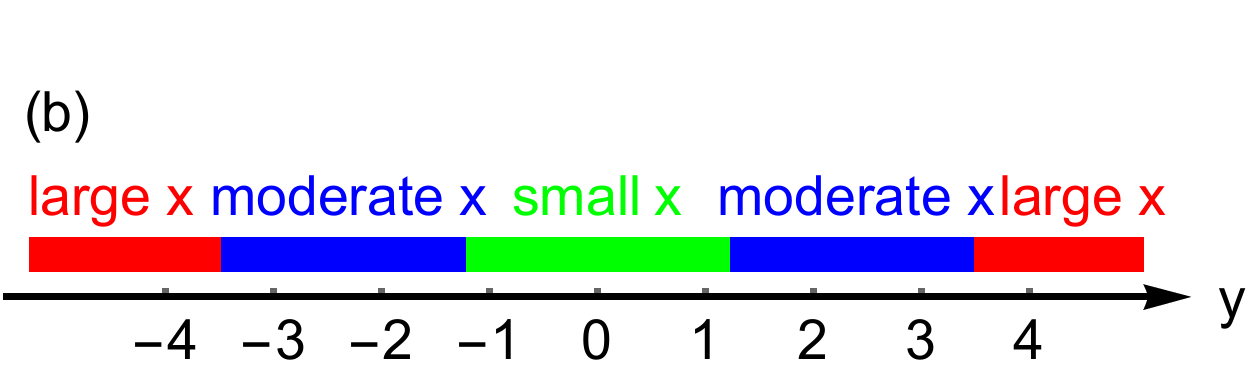}
\caption{(a) The quantum evolution of gluons in the right moving projectile. The black axis represents the longitudinal momentum fraction $x$ of partons in the projectile, and the blue axis roughly indicates corresponding rapidity $y$. (b) The quantitative correspondence between $x$ regions and rapidity regions for the case of $\mathrm{p_{\perp}}=$ 2 GeV$/$c in the collision of $\sqrt{s}=7$ TeV. }
\end{figure}

\begin{figure}[htbp]
\centering

\includegraphics[width=0.35\textwidth]{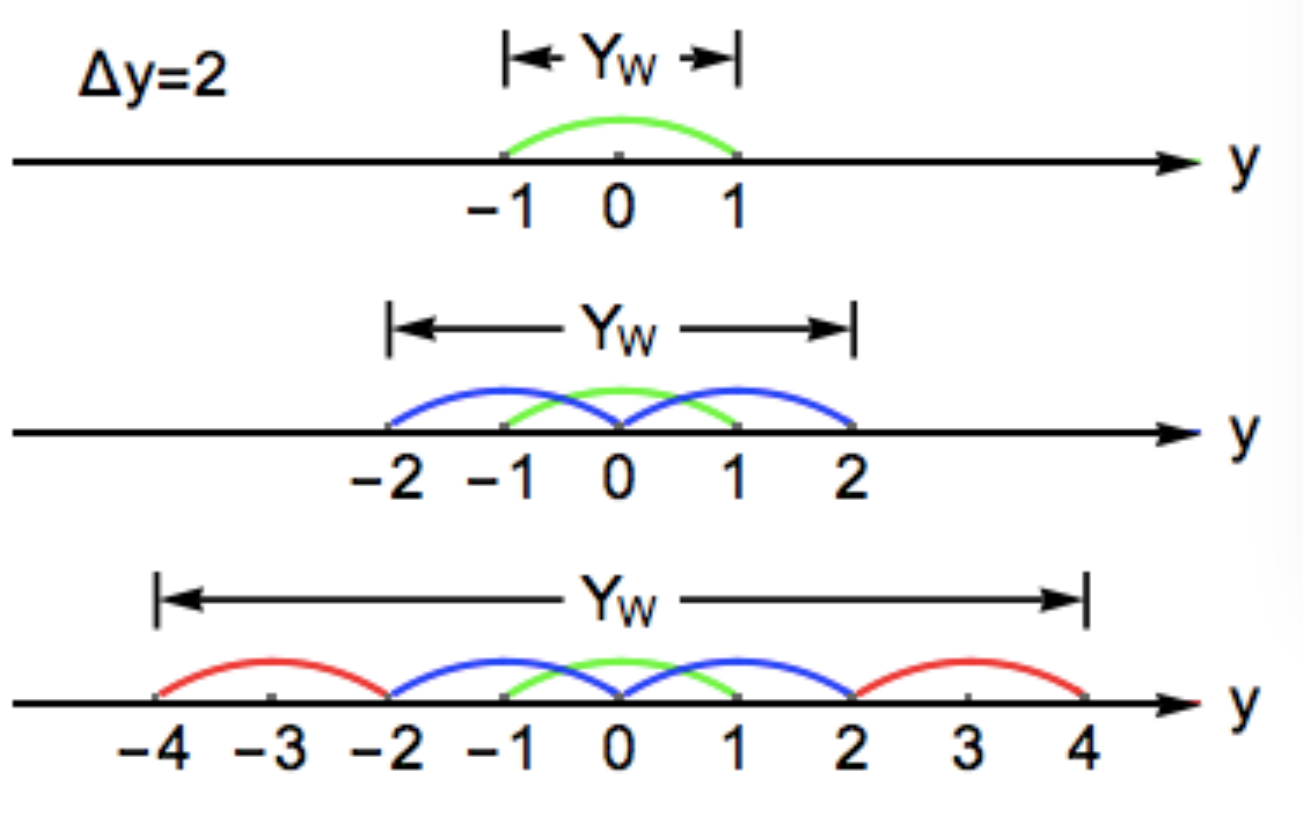}
\caption{A schematic plot for pairs with rapidity separation $\Delta y=2$. Rapidity window ($Y_{\rm W}$) of [-1, 1] only includes pairs with $y_{\rm p}=-1$ and $y_{\rm q}=1$, as the green curve shows. Rapidity window of [-2, 2] counts more pairs including pairs with $y_{\rm p}=0$ and $y_{\rm q}=\pm2$, as blue curves show. Rapidity window of [-4, 4] further includes pairs with $y_{\rm p}=\pm2$ and $y_{\rm q}=\pm4$, as red curves show. }
\label{Fig:2}
\end{figure}

\section{Two-gluon $\Delta y$-$\Delta\phi$ correlations from high energy QCD evolution}\label{sec:formula}

In a high energy collision, both the projectile and the target are regarded as high parton density sources. When they pass through each other, strong longitudinal color electric and magnetic fields are formed. The framework that describes the physics of high parton densities and strong color fields is the CGC effective field theory (CGC EFT)~\cite{Gribov-1983,Iancu-2003,Weigert-2005,Gelis-2010}. The effective degrees of freedom in this framework are color sources $\rho$ at large $x$ and classical gauge fields $\mathcal{A}_\mu$ at small $x$, as Fig.1(a) shows. The classical gauge field $\mathcal{A}_\mu$ can be obtained by numerically solving Yang-Mills equations with a given configuration of color source. For a given initial configuration of color source, fields in the nuclear wave functions evolve with Feynman $x$, which is described by the Jalilian-Marian-Iancu-McLerran-Weigert-Kovner (JIMWLK) renormalization group equations~\cite{Jalilian-Marian-1997,Jalilian-Marian-PRD,Iancu-2001}. In mean field approximation and large-$N_c$ limit, the JIMWLK equation is reduced to the Balitsky-Kovchegov (BK) equation~\cite{Balitsky1996,Balitsky1999,Kovchegov1999}, which describes the quantum evolution in Fig. 1(a).

Supposing two gluons are produced with transverse momentum $\mathbf{p}_\perp $, $\mathbf{q}_\perp$ and rapidity $y_{\rm p}$, $y_{\rm q}$, the correlation function is defined as,
\begin{widetext}
\begin{eqnarray}
C(\mathbf{p}_\bot, y_{\rm p}; \mathbf{q}_\bot, y_{\rm q}) &=& \frac{\frac{dN_2}{{d^{2}\mathbf{p}_\bot dy_{\rm p}}{d^{2}\mathbf{q}_\bot dy_{\rm q}}}}{\frac{dN_1}{d^{2}\mathbf{p}_\bot dy_{\rm p}}\frac{dN_1}{d^{2}\mathbf{q}_\bot dy_{\rm q}}}-1= \frac{\frac{dN_{\mathrm{2}}^{\rm corr}}{{d^{2}\mathbf{p}_\bot dy_{\rm p}}{d^{2}\mathbf{q}_\bot dy_{\rm q}}}}{\frac{dN_1}{d^{2}\mathbf{p}_\bot dy_{\rm p}}\frac{dN_1}{d^{2}\mathbf{q}_\bot dy_{\rm q}}},
\end{eqnarray}
\end{widetext}
where $\frac{dN_2}{{d^{2}\mathbf{p}_\bot dy_{\rm p}}{d^{2}\mathbf{q}_\bot dy_{\rm q}}}$, $\frac{dN_1}{d^{2}\mathbf{p}_\bot dy_{\rm p}}$ are the double- and single-gluon inclusive productions and $ \frac{dN_{2}^{\rm corr}}{{d^{2}\mathbf{p}_\bot dy_{\rm p}}{d^{2}\mathbf{q}_\bot dy_{\rm q}}}$ is the correlated double-gluon production. Subscripts ``p" and ``q" are used to mark the two gluons.

In the framework of CGC EFT, for a given collision, the observable under the leading-log approximation is factorized as~\cite{cgc-NPA-2008},
\begin{eqnarray}
\left\langle\mathcal{O}\right\rangle_{\mathrm{LLog}} &=& \int [D\rho_1][D\rho_2]W[\rho_1]W[\rho_2]\mathcal{O}[\rho_1,\rho_2]_{\mathrm{LO}},
\end{eqnarray}\\
where $\mathcal{O}[\rho_1,\rho_2]_{\mathrm{LO}}$ is leading-order single- or double- gluon inclusive distribution for a fixed distribution of color sources, and the integration denotes an average over different distributions of color sources with the weight functional $W[\rho_{1,2}]$. In general, $W[\rho_{1,2}]$ encodes all possible color charge configurations of the projectile and target, and obeys the JIMWLK renormalization group equations~\cite{Jalilian-Marian-1997,Jalilian-Marian-PRD,Iancu-2001}. All the quantum evolution of the projectile/target is absorbed into the distribution $W[\rho_{1,2}]$.

The averaging over color sources can be done under the McLerran-Venugopalan (MV) model with a Gaussian weight functional. According to Ref.~\cite{cgc-PRD-I,cgc-PRD-II,cgc-PRD-III}, the correlated two-gluon production can be expressed by unintegrated gluon distributions (uGD) as,
\begin{widetext}
\begin{eqnarray}
\frac{dN_{\mathrm{2}}^{\mathrm{corr}}}{{d^{2}\mathbf{p}_\bot dy_{\rm p}}{d^{2}\mathbf{q}_\bot dy_{\rm q}}}=\frac{C_2}{\mathbf{p}^2_\bot\mathbf{q}^2_\bot}
\left[\int \frac{d^2\mathbf{k}_\bot}{(2\pi)^2}(D_1+D_2)
+\sum_{j=\pm}\left[D_3(\mathbf{p}_\bot,j\mathbf{q}_\bot)+\frac{1}{2}D_4(\mathbf{p}_\bot,j\mathbf{q}_\bot)\right]\right],
\end{eqnarray}
\end{widetext}
where $C_2=\frac{\alpha_s(\rm{p}_\bot)\alpha_s(\rm{q}_\bot) N_c^2 S_{\bot}}{\pi^{8}(N_c^{2}-1)^{3}}$, and
\begin{widetext}\begin{eqnarray}
D_1 &=& \Phi_{A_1}^{2}(y_{\rm p},\mathbf{k}_\bot)\Phi_{A_2}(y_{\rm p},\mathbf{p}_\bot-\mathbf{k}_\bot)\left[\Phi_{A_2}(y_{\rm q},\mathbf{q}_\bot+\mathbf{k}_\bot)+\Phi_{A_2}(y_{\rm q},\mathbf{q}_\bot-\mathbf{k}_\bot) \right],   \nonumber \\
D_2 &=& \Phi_{A_2}^{2}(y_{\rm q},\mathbf{k}_\bot)\Phi_{A_1}(y_{\rm p},\mathbf{p}_\bot-\mathbf{k}_\bot)[\Phi_{A_1}(y_{\rm q},\mathbf{q}_\bot+\mathbf{k}_\bot)+\Phi_{A_1}(y_{\rm q},\mathbf{q}_\bot-\mathbf{k}_\bot)].
\end{eqnarray}\end{widetext}
Here, $\Phi_{A_{1(2)}}(y,\mathbf{k}_\bot)$ denotes the uGD of projectile $A_1$ or target $A_2$. In Eq.(4), 
\begin{equation}
D_3(\mathbf{p}_\bot, j\mathbf{q}_\bot)=\delta^{2}(\mathbf{p}_\bot+j\mathbf{q}_\bot)\left[\mathcal{I}_1^{2}+\mathcal{I}_2^{2}+2\mathcal{I}_3^{2}\right],
\end{equation}
with
\begin{widetext}
\begin{eqnarray}
\mathcal{I}_1   &=&  \int \frac{d^2\mathbf{k}_{1\bot}}{(2\pi)^2} \Phi_{A1}(y_{\rm p},\mathbf{k}_{1\bot})\Phi_{A2}(y_{\rm q},\mathbf{p}_\bot-\mathbf{k}_{1\bot})\frac{(\mathbf{k}_{1\bot} \cdot \mathbf{p}_{\bot}-\mathbf{k}_{1\bot}^{2})^{2}}
{\mathbf{k}_{1\bot}^{2}(\mathbf{p}_{\bot}-\mathbf{k}_{1\bot})^{2}}, \\
\mathcal{I}_2   &=&  \int \frac{d^2\mathbf{k}_{1\bot}}{(2\pi)^2} \Phi_{A1}(y_{\rm p},\mathbf{k}_{1\bot})\Phi_{A2}(y_{\rm q},\mathbf{p}_\bot-\mathbf{k}_{1\bot})\frac{|\mathbf{k}_{1\bot}\times\mathbf{p}_{\bot}|^{2}}{\mathbf{k}_{1\bot}^{2}
(\mathbf{p}_{\bot}-\mathbf{k}_{1\bot})^{2}}, \\
\mathcal{I}_3   &=&  \int \frac{d^2\mathbf{k}_{1\bot}}{(2\pi)^2} \Phi_{A1}(y_{\rm p},\mathbf{k}_{1\bot})\Phi_{A2}(y_{\rm q},\mathbf{p}_\bot-\mathbf{k}_{1\bot})\frac{(\mathbf{k}_{1\bot} \cdot \mathbf{p}_{\bot}-\mathbf{k}_{1\bot}^{2})|\mathbf{k}_{1\bot}
\times\mathbf{p}_{\bot}|}{\mathbf{k}_{1\bot}^{2}(\mathbf{p}_{\bot}-\mathbf{k}_{1\bot})^{2}},
\end{eqnarray}
\end{widetext}
and
\begin{widetext}\begin{eqnarray}
D_4(\mathbf{p}_{\bot},j\mathbf{q}_{\bot})   &=&  \int \frac{d^2\mathbf{k}_{1\bot}}{(2\pi)^2} \Phi_{A1}(y_{\rm p},\mathbf{k}_{1\bot})\Phi_{A1}(y_{\rm p},\mathbf{k}_{2\bot})\Phi_{A2}(y_{\rm q},\mathbf{p}_\bot-\mathbf{k}_{1\bot})\Phi_{A2}
(y_{\rm q},\mathbf{p}_\bot-\mathbf{k}_{2\bot})  \nonumber \\
   &\times&  \frac{(\mathbf{k}_{1\bot} \cdot \mathbf{p}_{\bot}-\mathbf{k}_{1\bot}^{2}){(\mathbf{k}_{2\bot} \cdot \mathbf{p}_{\bot}-\mathbf{k}_{2\bot}^{2})}+(\mathbf{k}_{1\bot}\times\mathbf{p}_{\bot}) \cdot (\mathbf{k}_{2\bot}
   \times\mathbf{p}_{\bot})}{\mathbf{k}_{1\bot}^{2}(\mathbf{p}_{\bot}-\mathbf{k}_{1\bot})^{2}}   \nonumber \\
   &\times& \frac{(\mathbf{k}_{1\bot} \cdot j\mathbf{q}_{\bot}-\mathbf{k}_{1\bot}^{2}){(\mathbf{k}_{2\bot} \cdot j\mathbf{q}_{\bot}-\mathbf{k}_{2\bot}^{2})}+(\mathbf{k}_{1\bot}\times\mathbf{q}_{\bot}) \cdot (\mathbf{k}_{2\bot}
   \times\mathbf{q}_{\bot})}{\mathbf{k}_{2\bot}^{2}(j\mathbf{q}_{\bot}-\mathbf{k}_{1\bot})^{2}},
\end{eqnarray}\end{widetext}
where $\mathbf{k}_{2\bot}\equiv\mathbf{p}_{\bot}-\mathbf{k}_{1\bot}+j\mathbf{q}_{\bot}$.

The single-gluon inclusive production reads
\begin{eqnarray}
&&\frac{dN_1}{d^{2}\mathbf{p}_\bot dy_{\rm p}} \nonumber \\ &=& \frac{\alpha_s({\rm p}_\bot) N_c S_{\bot}}{\pi^{4}(N_c^{2}-1)}\frac{1}{\mathbf{p}_\bot^2}\int\frac{d\mathbf{k}_\bot^2}{(2\pi)^2}\Phi_{A}(y_{\rm p},\mathbf{k}_\bot)\Phi_{A}
(y_{\rm p},\mathbf{p}_\bot-\mathbf{k}_\bot).\nonumber\\
\end{eqnarray}

Based on correlation function $C(\mathbf{p}_\bot, y_{\rm p}; \mathbf{q}_\bot, y_{\rm q})$, the associated yield per trigger is defined as
\begin{eqnarray}
Y(\Delta \phi,\Delta y) = \frac{1}{N_{\rm{Trig}}}\frac{d^{2}N_{\rm{Assoc}}}{d\Delta \phi d\Delta y},
\end{eqnarray}
where
\begin{widetext}
\begin{eqnarray}
\frac{d^{2}N_{\rm{Assoc}}}{d\Delta \phi d\Delta y} &=&\int_{y^{\rm min}}^{y^{\rm max}} dy_{\rm p}\int_{y^{\rm min}}^{y^{\rm max}} dy_{\rm q} \delta(y_{\rm q}-y_{\rm p}-\Delta y)\int_0^{2\pi} d\phi_{\rm p} \int_0^{2\pi} d\phi_{\rm q}  \delta(\phi_{\rm q}-\phi_{\rm p}-\Delta \phi) \nonumber\\ &&\times\int^{{\rm p}^{\rm{max}}_\bot}_{{\rm p}^{\rm{min}}_\bot}\frac{d{\rm p}^{2}_\bot}{2}\int^{{\rm q}^{\rm{max}}_\bot}_{{\rm q}^{\rm{min}}_\bot}\frac{d{\rm q}^{2}_\bot}{2}\frac{dN^{\rm{corr}}_{\mathrm{2}}}{{d^{2}\mathbf{p}_\bot dy_{\rm p}}{d^{2}\mathbf{q}_\bot dy_{\rm q}}},
\end{eqnarray}\end{widetext}
and
\begin{eqnarray}
N_{\rm{Trig}} &=& \iiint_{\mathrm{Acceptance}}dy d^{2}\mathbf{p}_\bot\frac{dN_1}{d^{2}\mathbf{p}_\bot dy_{\rm p}}.
\end{eqnarray}

The associated yield per trigger $Y(\Delta \phi,\Delta y)$ describes correlations of two gluons with rapidity separation $\Delta y$ and azimuthal separation $ \Delta\phi$ in given transverse momentum intervals ($\mathrm{p^{min}_{\bot}}$, $\mathrm{p^{max}_{\bot}}$), ($\mathrm{q^{min}_{\bot}}$, $\mathrm{q^{max}_{\bot}}$) and rapidity window $(y^{\rm min},y^{\rm max})$. In the following we do not distinguish between trigger gluons and associated gluons, but we still use the name ``associated yield per trigger" for simplicity.
 
The framework is valid to leading-logarithmic accuracy in $x$ and momentum ${\rm p}_\bot$, ${\rm q}_\bot \mathrm{\gtrsim} Q_s$. The important ingredient in the above expressions is uGD ($\Phi$), which can be obtained by solving the BK equation with running coupling corrections with a given initial condition. To avoid repetition, details can be found in Ref. \cite{cgc-NPA-2010} and our previous paper \cite{zhao-1,zhao-2}. For pp collision at $7\ \mathrm{TeV}$, $ Q^2_{s0}$ (with $Q_{s0}$ the initial value of $Q_s$ at $x_0$) is chosen to be $0.168\ \mathrm{GeV^2}$ \cite{cgc-PRD-I}.

The solution of the BK equation is reliable when $x\leqslant 0.01$. For larger values of $x$, the BK description breaks down. A phenomenological extrapolation for the unintegrated gluon distribution has the form
\begin{equation}
\Phi(x,\mathbf{k}_\bot)=(\frac{1-x}{1-x_0})^\beta\Phi(x_0,\mathbf{k}_\bot), \mbox{ for } x>x_0,
\end{equation}
where $x_0=0.01$ and the parameter $\beta=4$\cite{cgc-NPA-2010}. 

Both the experimental data~\cite{CMS-pp-JHEP-2010,ALICE-pPb-PLB-2016} and the CGC~\cite{cgc-PRD-I,cgc-PRD-II,cgc-PRD-III} show that the correlation at near side gets strongest only in an intermediate $\mathrm{p_\bot}$ interval, approximately $1 < \mathrm{p_\bot} < 3\ \mathrm{GeV/c}$. A calculation from the CGC points out that the correlation function gets a maximum when the transverse momenta are close to $\mathrm{2Q_{sp}}\sim 2\ \mathrm{GeV/c}$ for minimum-bias pp collisions, where $\mathrm{Q_{sp}}$ denotes the saturation momentum of proton \cite{zhao-1}. To obtain the strongest correlation, the correlation function is integrated within $1\leq \mathrm{p_\bot}(\mathrm{q_\bot}) \leq3\ \mathrm{GeV/c}$ here.  

\section{Rapidity window dependence of the ridge correlations}\label{sec:results}

\begin{figure*}[htb]
\begin{centering}
\begin{tabular}{ccc}
$\vcenter{\hbox{\includegraphics[scale=0.29]{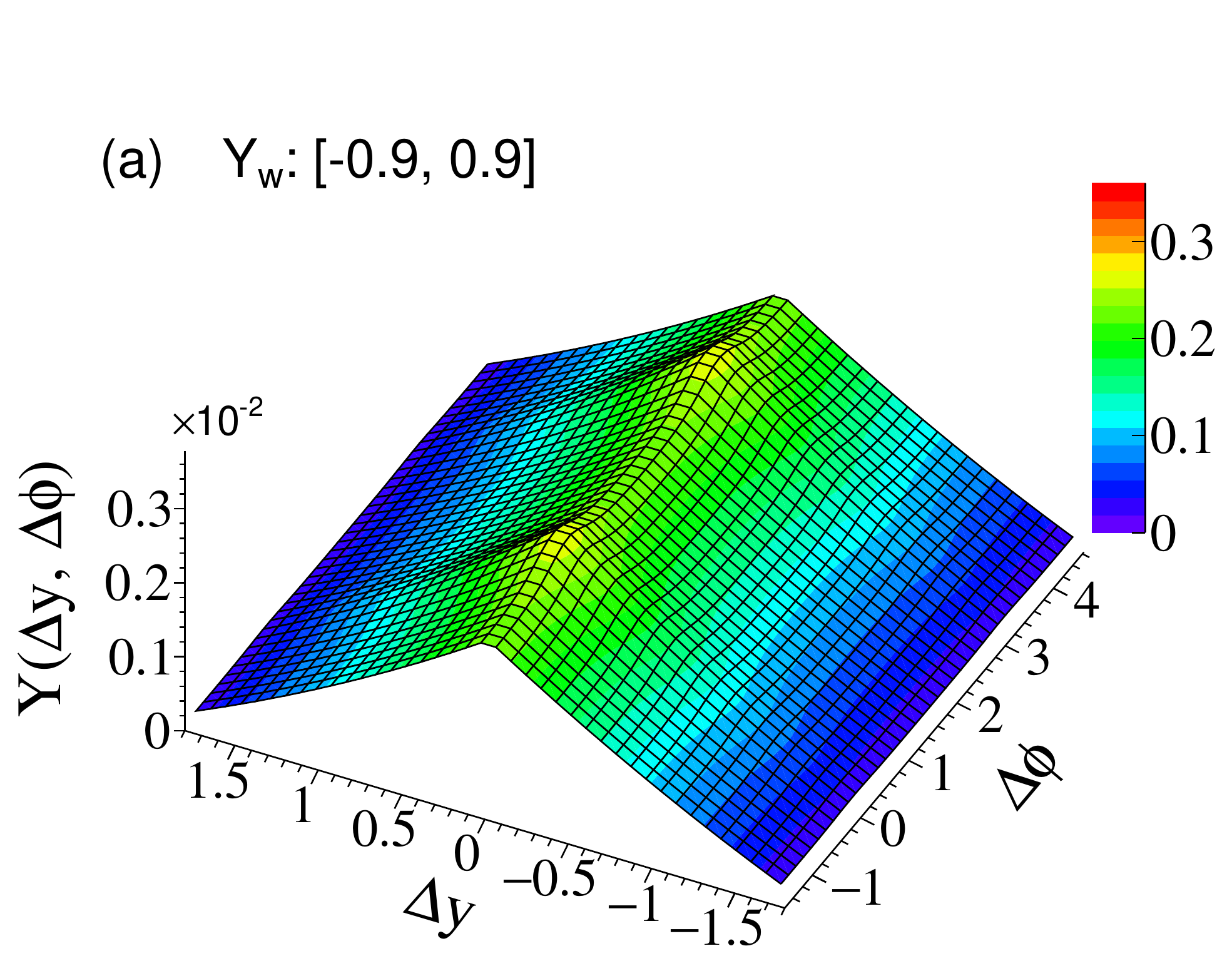}}}$ &
$\vcenter{\hbox{\includegraphics[scale=0.29]{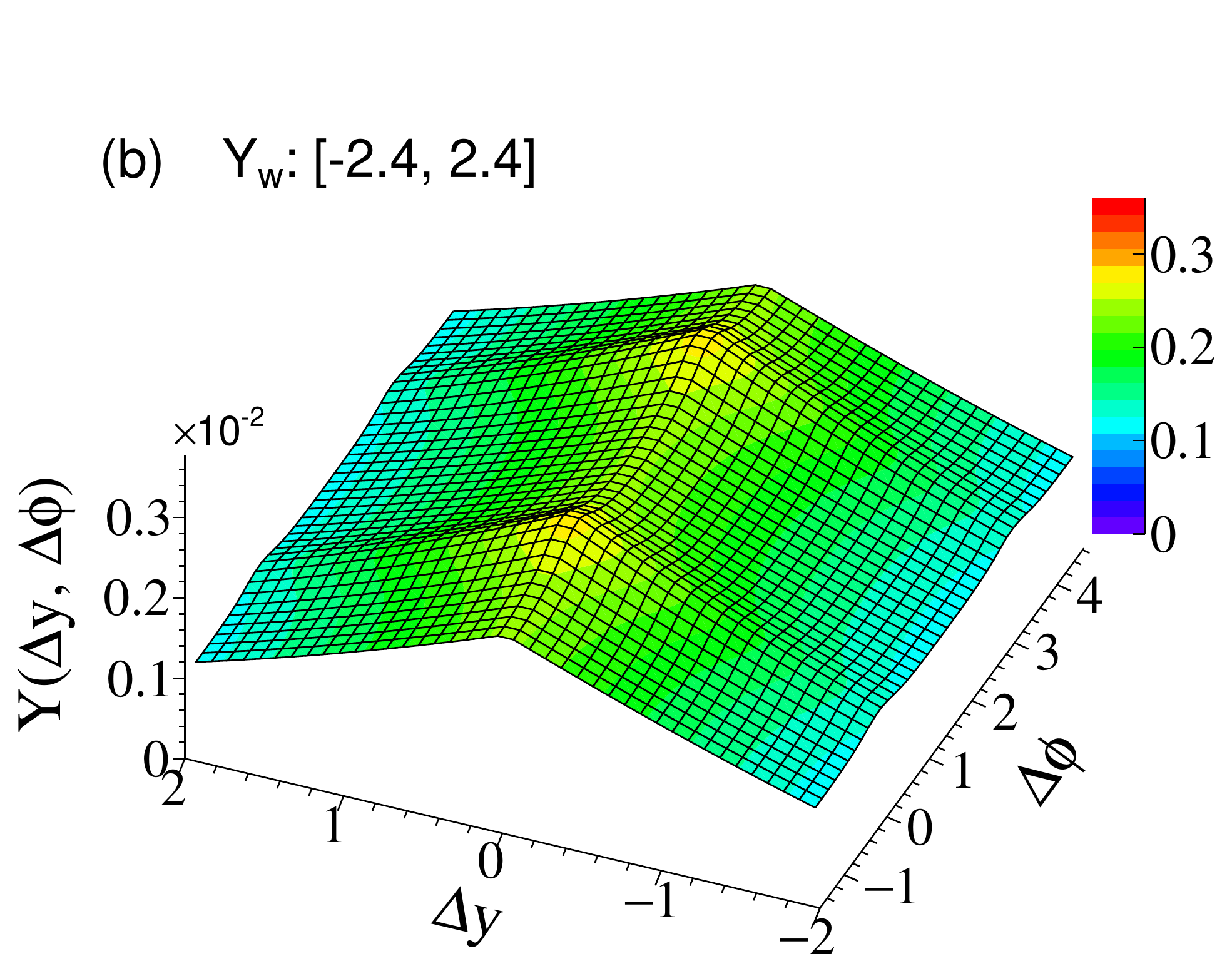}}}$
& $\vcenter{\hbox{\includegraphics[scale=0.29]{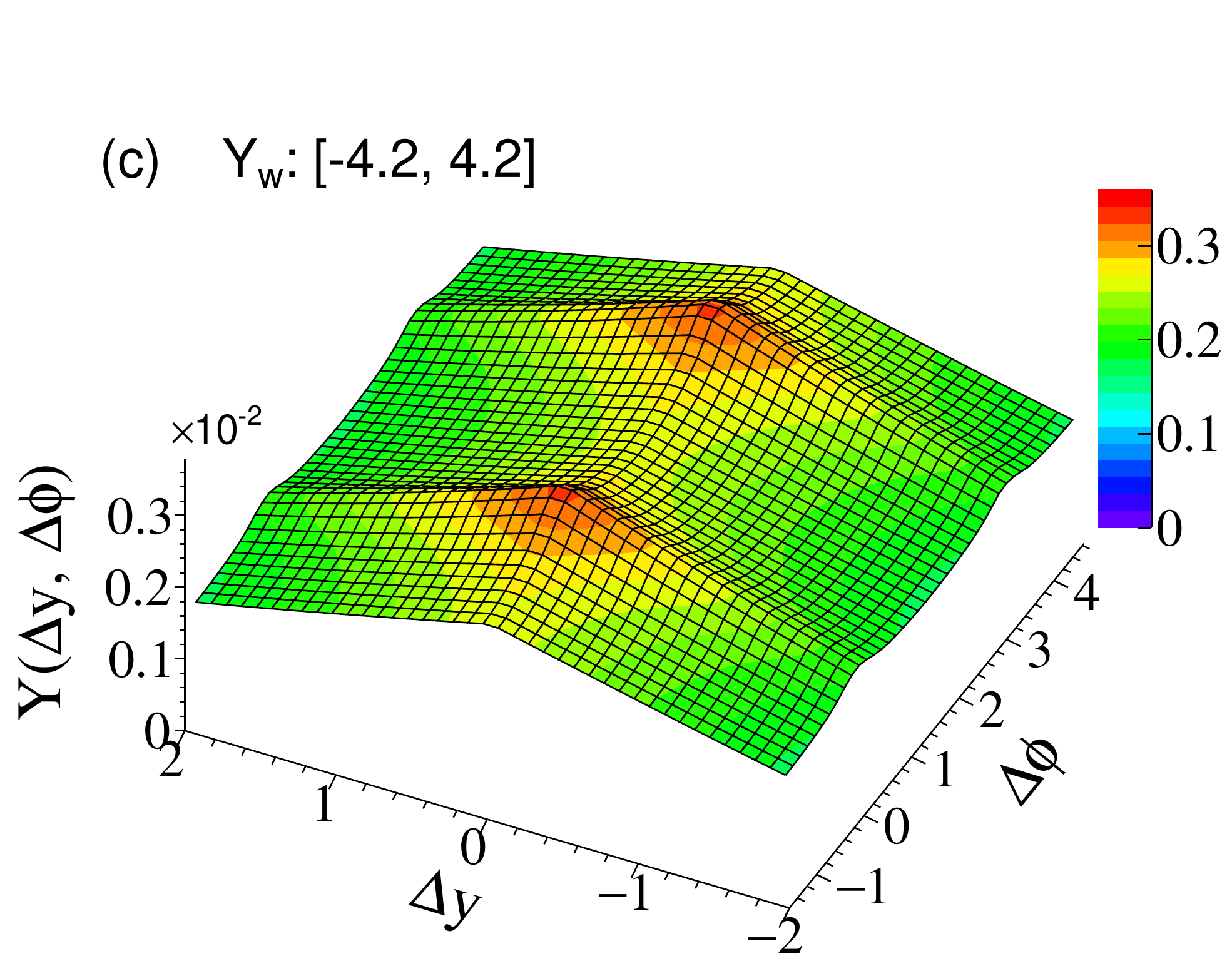}}}$ 
\end{tabular}
\par\end{centering}
\caption{The associated yield per trigger on $\Delta y$-$\Delta\phi$ plane for 7 TeV pp collisions with transverse momentum integrated within $1\leq p_{\perp}(q_{\perp})\leq 3$GeV$/$c and rapidity integrated in (a) $-0.9\leq y_{\rm p}(y_{\rm q})\leq 0.9$, (b) $-2.4\leq y_{\rm p}(y_{\rm q})\leq2.4$ and (c) $-4.2\leq y_{\rm p}(y_{\rm q})\leq 4.2$, respectively.}
\label{y-3-window}
\end{figure*}

Ridge correlations are usually illustrated by two dimensional $\Delta y$-$\Delta\phi$ distribution and one dimensional $\Delta\phi$ distribution. In this paper we start with the calculation of the $\Delta y$-$\Delta\phi$ distribution, i.e. the associated yield per trigger in Eq.~(12), which is widely measured in CMS, ALICE and ATLAS experiments. 
 
In order to study the rapidity window dependence of the long range ridge correlations, three rapidity windows, i.e. $[-0.9, 0.9], [-2.4, 2.4]$ and $[-4.2, 4.2]$, are used. The correlations in pp collisions at $\sqrt{s}$ = 7 TeV with transverse momentum integrated within $1\leq \mathrm{p_\bot}(\mathrm{q_\bot}) \leq3\ \mathrm{GeV/c}$ are plotted on  $\Delta y$-$\Delta\phi$ plane in Fig.~3.   

The rapidity window $[-0.9, 0.9]$ is chosen to be the same with the ALICE acceptance. As Fig.~3(a) shows, the correlations have two moderate peaks of equal height at $\Delta\phi=0$ and $\pi$ for a fixed $\Delta y$. For a fixed  $\Delta\phi$, the correlations show a downward trend as $|\Delta y|$ increases.   

As rapidity window extends to $[-2.4, 2.4]$, i.e. the CMS acceptance, the peaks at $\Delta\phi=0$ and $\pi$ get pronounced. The double ridges, i.e. two raised lines along $\Delta y$ direction at $\Delta\phi=0$ and $\pi$ on the surface plot, are clearly seen in Fig.~3(b). The range of $\Delta y$ is only plotted to $\pm2$ so that a direct comparison can be made with Fig.~3(a). For longer range ($|\Delta y|>2$) the ridges extend continuously which are not shown in the plot.  

When rapidity window extends to $[-4.2, 4.2]$, the surface in Fig.~3(c) is overall uplifted compared to Fig.~3(a) and 3(b). The  correlations at long range rapidity are uplifted more, as the color at  large $|\Delta y|$ changes from blue in Fig.~3(a) to light blue in Fig.~3(b) to green in Fig.~3(c). When it comes to the $\Delta\phi$ direction, the amplitude of $\Delta\phi$ distribution, both at short range and long range rapidity gap, increases significantly with increasing rapidity window. 

Fig.~3 demonstrates an enhancement of correlations at long range rapidity gap for larger $Y_{\rm W}$. As $Y_{\rm W}$ increases, rapidity correlations at fixed $\Delta\phi$ change from a steep to a relatively flat shape, indicating that long range rapidity correlations increase with rapidity window. On the other hand, the amplitude of $\Delta\phi$ distribution at fixed $\Delta y$ also increases with rapidity window.    

The downward trend in $\Delta y$ direction is gentle in the ALICE measurement of the associated yield per trigger in pPb collisions at a center-of-mass energy of 5.02 TeV~\cite{ALICE-pPb-PLB-2013}. Compared to the ALICE results, the downward trend here is much more steep. When rapidity window extends to $[-2.4, 2.4]$, ridges that can extend to $|\Delta y|=4$ are not as flat as those obtained in CMS measurements, either~\cite{CMS-pp-JHEP-2010}. It shows that glasma graphs have significant short range rapidity correlations~\cite{zhao-2}.

Of particular interest in studies of ridge is the long range (large $|\Delta y|$) structure of two-particle correlation functions.  In order to get a better view of the ridge correlation at long range rapidity and its enhancement, we plot one dimensional $\Delta\phi$ distribution at fixed long range rapidity gap in Fig.~4(a). For $Y_{\rm W}$ of $[-0.9, 0.9]$, the $\Delta\phi$ distribution at $\Delta y=1.8$ (1.8 units are the maximum rapidity gap for that rapidity window) is straight, as the black curve shows. In fact, the two peaks are still there. They are not visible in the plot just because the amplitude is small. For  $Y_{\rm W}$ of $[-2.4, 2.4]$, the $\Delta\phi$ distribution at $\Delta y=2$ has a larger amplitude and the peaks at 0 and $\pi$ is clearly seen, as the blue curve shows. As rapidity window gets larger to $[-4.2, 4.2]$, the peaks get more pronounced and the amplitude of the distribution grows larger, as the red curve shows. 

\begin{figure*}[htb]
\begin{centering}
\begin{tabular}{cc}
$\vcenter{\hbox{\includegraphics[scale=0.4]{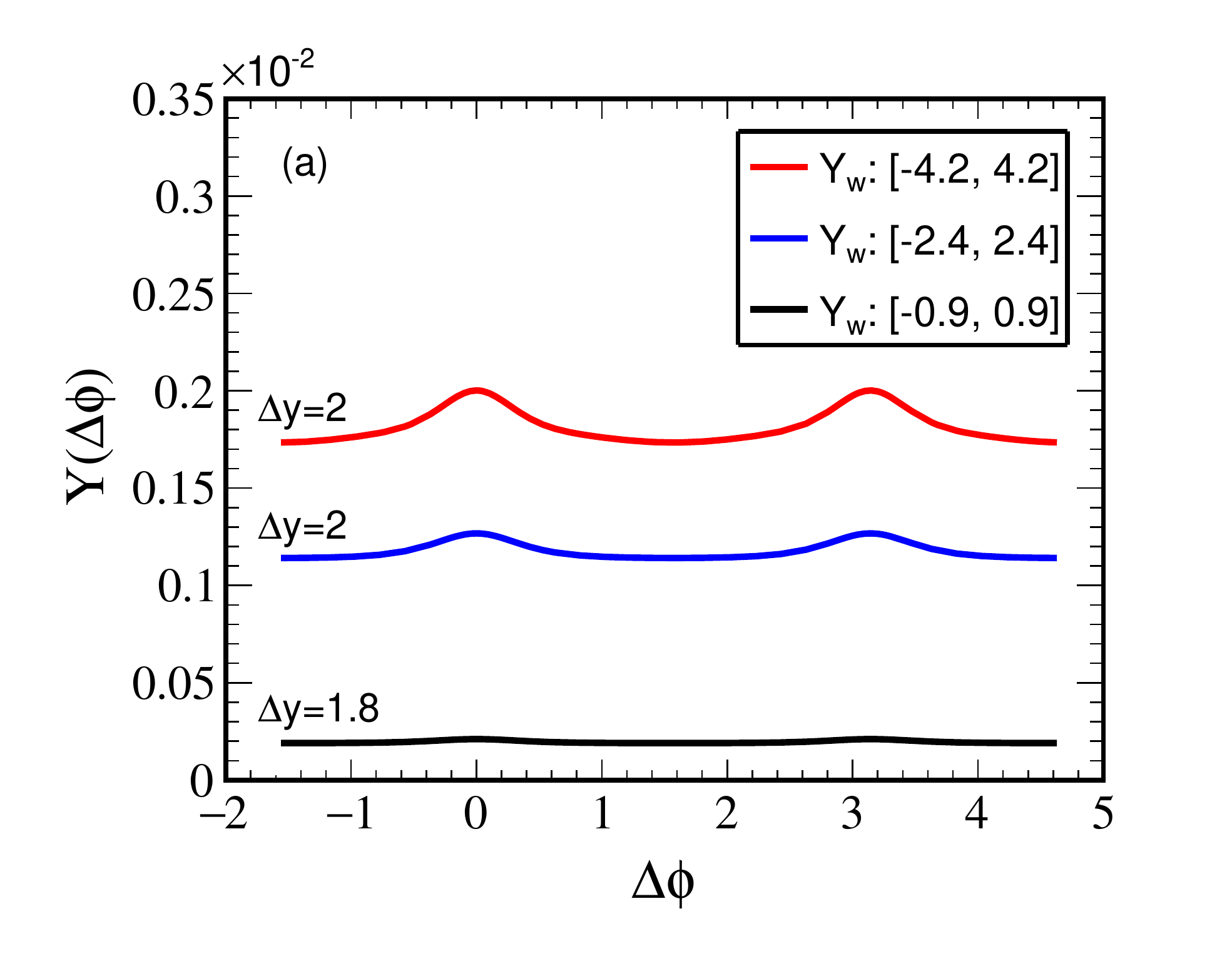}}}$ &
$\vcenter{\hbox{\includegraphics[scale=0.4]{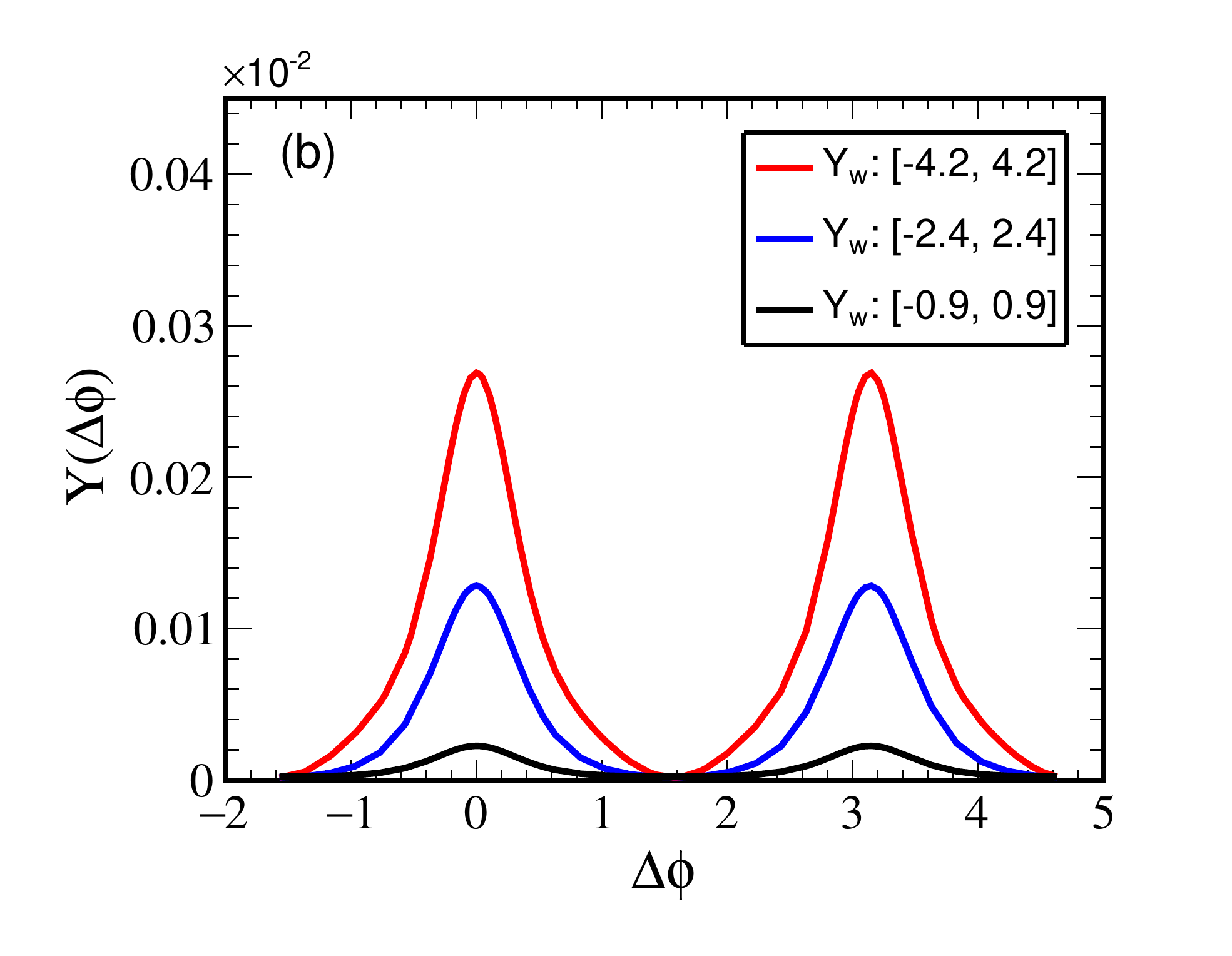}}}$ 
\end{tabular}
\par\end{centering}
\caption{(a) The $\Delta\phi$ distribution at a fixed long range rapidity gap. (b) The $\Delta\phi$ distribution with a constant background subtracted by the ZYAM method. }
\end{figure*}

According to the zero-yield-at-minimum (ZYAM) method~\cite{note}, the integrated associated yield is defined as the area under the peak of the $\Delta\phi$ distribution above a constant background. The $\Delta\phi$ distribution with a constant background subtracted by the ZYAM method is shown in Fig.~4(b). ZYAM subtraction scheme makes the signals cleaner. The biggest amplitude is the red curve, corresponding to the widest rapidity window. The area under the peaks for the three curves qualitatively shows that the integrated associated yield grows drastically with rapidity window. 

The two peak structure of $\Delta\phi$ distribution at long range rapidity for the three rapidity windows confirms an intrinsic collimation production in the CGC formalism. Quantitative calculations about the near side ridge yield in pp and pPb collisions are consistent with CMS, ATLAS and ALICE data~\cite{cgc-PRD-I,cgc-PRD-II,cgc-PRD-III}. It means that small $x$ dynamics can explain the ridge yield. The enhancement at wider rapidity window, demonstrated in Fig.~4(b) of this paper, indicates that large $x$ gluons, i.e. source gluons, make significant contributions to the ridge yield. 

The ridge structure is twofold. The azimuthal collimation in CGC comes from Cauchy-Schwartz inequality~\cite{Dumitru-PLB-2011}, which states that the azimuthal correlations get maximum when the two transverse momenta are parallel or antiparallel. The ridge at long range rapidity is due to the contribution of color sources. Beyond that, we also demonstrate that source gluons enhance the azimuthal collimation, indicating an interplay between longitudinal and transverse directions.


The rapidity window of track reconstruction from the ALICE detector is very different from the CMS and ATLAS. Whether the data from the three experiments at the LHC shows the increasing trend with rapidity window as shown in Fig.~4(b) is hard to conclude at present. As mentioned before, long range angular correlations are sensitive to centrality class and $\mathrm{p_\bot}$ window. In this paper we show that long range angular correlations are also sensitive to rapidity window. In order to see the trend with rapidity window, the correlations should be compared at the same centrality and the same $\mathrm{p_\bot}$ window. The ALICE collaboration reports long range angular correlations in a centrality bin that is different from CMS and ATLAS~\cite{CMS-pPb-PLB-2013,ALICE-pPb-PLB-2013, ATLAS-pPb-PRL-2013}, which makes a direct comparison between the three experiments impossible at the moment.

Enlarging the rapidity window is one way of seeing the contributions of source gluons. Obviously, the integration over a given rapidity window includes different combinations of gluons. In order to further clarify the contributions of source gluons, a more direct observable is the differential correlation function $C(\mathbf{p}_\bot, y_{\rm p}; \mathbf{q}_\bot, y_{\rm q})$, i.e. Eq.(2) with specific $y_{\rm p}$ and $y_{\rm q}$, which is shown in Fig.5. 


\begin{figure}[htb]
\centering
\includegraphics[width=0.45\textwidth]{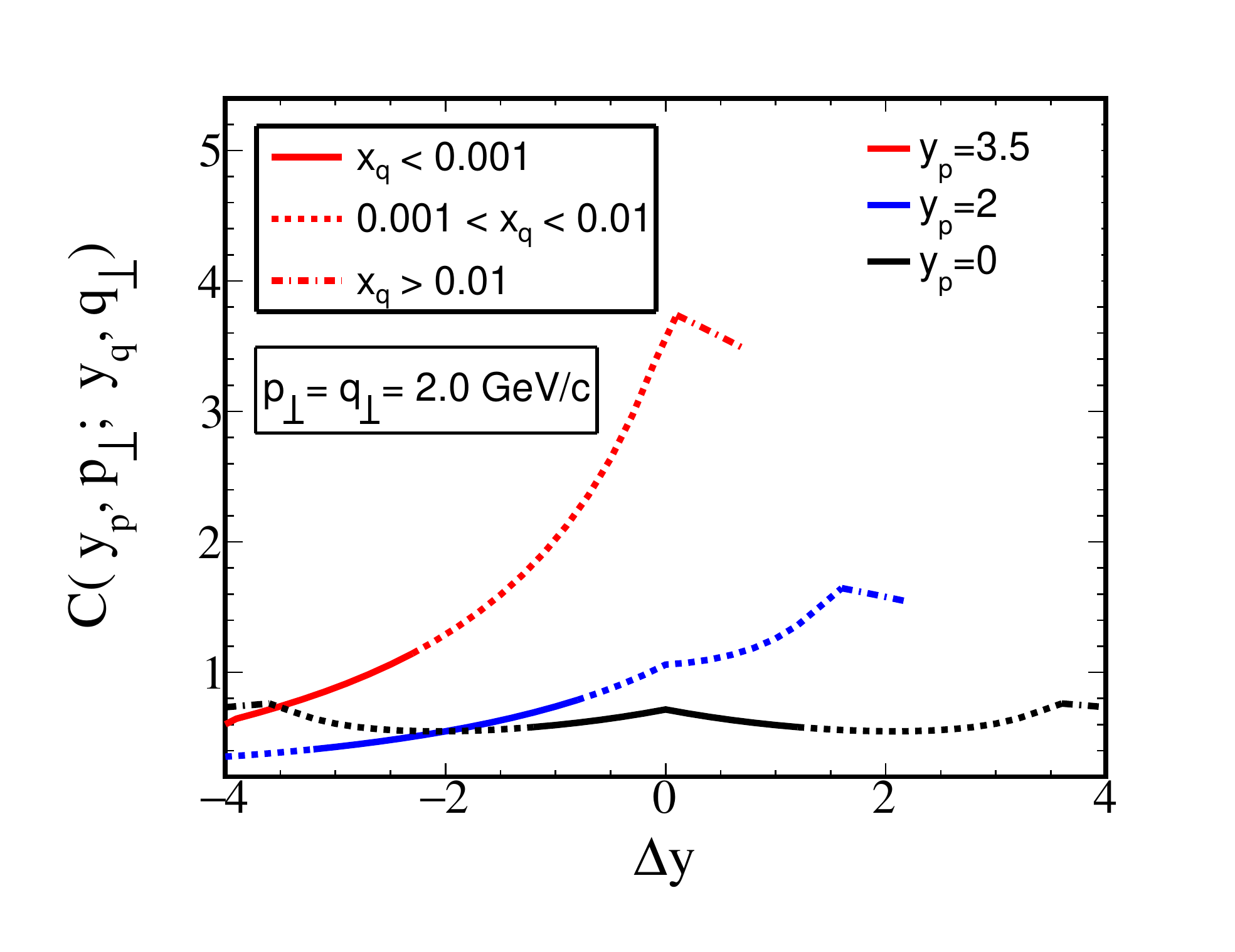}
\caption{The differential correlation function at $\mathrm{p_\bot}=\mathrm{q_\bot}=2.0$ GeV/c with $\phi_{\rm p}=\phi_{\rm q}=0$. Black, blue and red curves denote $y_{\rm p}=0$, 2 and 3.5, respectively. }
\label{Fig:5}
\end{figure}

A similar figure has already been shown in ref~\cite{zhao-1}. The motivation of presenting it here again is to understand the integrated correlations by the differential correlations. The enhanced correlations originating from source gluons is reflected more directly by the rapidity location dependence of the differential correlation function. 

The differential correlation function in Fig.~5 shows some interesting correlation patterns. In the case of $y_{\rm p}=0$, i.e. the trigger particle is at small $x$, the correlation function is almost flat and shows a rising trend when $\Delta y$ exceeds 3 as the black curve shows. The solid, dash and dash-point lines denote $x_{\rm q}<10^{-3}$, $10^{-3}<x_{\rm q}<10^{-2}$ and $x_{\rm q}>10^{-2}$, labeling the results of the small, moderate and large-$x$ degrees of freedom of the associated particle, respectively. The point connecting the dash line and dash-point line represents $x_0=0.01$. Since quantitative results of the extrapolation Eq.~(15) is unreliable, we only take the dash-point lines that are near the connection points as large-$x_{\rm q}$. 

A rapidity difference of only 2 units is sensitive to the small $x$ evolution, denoted by the solid lines in Fig.~5. A wider region is sensitive to the moderate and large $x$ degrees of freedom. The black curve indicates that the correlation strength between radiated gluons (r for short), moderate-$x$ gluons (m for short) and source gluons (s for short) has 
\begin{equation}C_{\rm rr} \approx C_{\rm rm} < C_{\rm rs},\end{equation} where each subscript represents a kind of gluon. It shows that correlations between radiated gluons are not as strong as those with source gluons.

Similarly, the rising trend of the blue curve in Fig.~5 indicates \begin{equation}C_{\rm mr} < C_{\rm mm} < C_{\rm ms},\end{equation} and the sharp rise of  the red curve indicates \begin{equation}C_{\rm sr} \ll C_{\rm sm} \ll C_{\rm ss}.\end{equation} $C_{\rm rr}$, $C_{\rm rm}$ and $C_{\rm ms}$ represents the green, blue and red curves in Fig.~2, respectively. The strong correlations between the moderate-$x$ gluon and the source gluon, i.e. $C_{\rm ms}$, make the ridge yield enhanced significantly, as the red curve in Fig.~4(b) shows.

The ordering in Eqs. (16)-(18) demonstrates that correlations with source gluons are the strongest for each type of trigger gluon. For trigger gluons at non-central rapidity, like the case of $y_{\rm p}=2$ and 3.5, correlations with small-x, moderate-x and large-x gluons show a continuously increasing trend. The strongest correlations exist between source gluons. These ordering patterns, especially the continuously rising trend of correlations between trigger particle at non-central rapidity and associated particle at other rapidities demonstrate characteristic features of CGC. 

\section{Summary and discussion}\label{sec:summary}

According to the relation of Feynman $x$ to rapidity $y$, different rapidity regions reveal gluon dynamics at different $x$ and thus the rapidity window dependence of long range ridge correlations is sensitive to the quantum evolution of gluon saturation dynamics.   

The ridge structure is twofold, i.e. the longitudinal and transverse structures. Two dimensional $\Delta y$-$\Delta\phi$ distributions illustrate that ridge structures in both directions have rapidity window dependence. As the rapidity window enlarges, the longitudinal structure has a rising and flattening trend at long range rapidity gap. It indicates an enhancement of long range rapidity correlations. On the other hand, the angular distribution at long range rapidity gap shows larger amplitudes as rapidity window enlarges. It indicates the enhancement of azimuthal correlations and the resulting ridge yields. In summary, ridge correlations in both longitudinal and transverse directions get stronger as the rapidity window enlarges. 

Within the CGC framework, the azimuthal collimation comes from Cauchy-Schwartz inequality which requires parallel or antiparallel of two gluon transverse momenta to get the maximum correlations. The long range rapidity correlations is due to the contribution of color sources. As the rapidity window enlarges, source gluons are included and begin to play a role. The enhancement in both longitudinal and transverse correlations demonstrates that source gluons enhance the long range rapidity correlations and azimuthal collimation.


While the origin of the ridge in small systems is inconclusive, it is already widely known that the color flux tube picture naturally explains the long range rapidity correlation and angular collimation within the CGC framework. This paper notices that the rapidity of gluons is related to the quantum evolution and the inclusion of large rapidity, or equivalently the source gluons, will enhance the long range rapidity ridge correlations significantly.

So far the correlation patterns calculated are for gluons. It is shown in Ref.~\cite{Dusling-PRL-2012} that fragmentation functions only have a major impact on transverse momentum dependence. The correlation patterns as a function of rapidities would remain in the final state. 


Since the quantum evolution of the CGC gives characteristic rapidity location dependence of the differential correlation, differential measurements are more direct for probing the CGC and are highly needed in the future measurement. The correlation patterns of one particle at non-central rapidities with the other particle at other rapidities can provide sensitive tests of this picture. Similar suggestions have been made in Ref.~\cite{Dumitru-PLB-2011}.  

The work of Ref.\cite{Schenke-PRL-2016} where IP-glasma model is combined with Lund string fragmentation demonstrates similar ridge structures. Flatter ridges along rapidity direction shown there are a result of an exactly boost invariant distribution implemented. The similarity in ridge structures indicates that the long range rapidity ridge correlation in high energy hadronic collisions is intrinsic to glasma dynamics.

Ultra-long range rapidity ridge, up to nearly 8 units of pseudorapidity separation, has been observed in a preliminary result of ALICE experiment~\cite{ultra-long}. The picture of classical fields has a restricted range of validity. In a larger range of rapidity, quantum fluctuations become important, and the description in terms of classical fields breaks down. For large rapidity, one has less theoretical control within this framework. The contribution of quarks should be considered.

\section*{Acknowledgement}
This work is supported in part by the NSFC of China under Grant No. U1732271. 

\providecommand{\href}[2]{#2}\begingroup\raggedright\endgroup

\begin{thebibliography}{10}
\bibitem{note}An implementation of the zero-yield-at-minimum (ZYAM) method firstly calculates the integral of $Y(\Delta y,\Delta\phi)$ over $2.0<|\Delta y|<4.8$ to give $R(\Delta\phi)$. Then a second-order polynomial is fitted to $R(\Delta\phi)$ in the region $0.5<|\Delta\phi|<1.5$. The location of the minimum of the polynomial in this region is denoted $\Delta\phi_{\rm ZYAM}$. Using the position of the minimum, the integrated associated yield is then calculated as the integral of $R(\Delta\phi)$ minus its value at $\Delta\phi_{\rm ZYAM}$ over the region $0<|\Delta\phi|<\Delta\phi_{\rm ZYAM}$.
\end{thebibliography}
\end{document}